\documentclass[english,aps,prd,preprintnumbers, nofootinbib,
superscriptaddress,showkeys,showpacs,byrevtex,fleqn]{revtex4-1}
\usepackage[utf8]{inputenc}
\usepackage{color}
\usepackage{array}
\usepackage{bm}
\usepackage{multirow} 
\usepackage{amsmath}
\usepackage{amssymb}
\usepackage{amsfonts}
\usepackage{amscd}
\usepackage{amsxtra}
\usepackage{amsthm}
\usepackage{babel}
\usepackage{graphicx}

\makeatletter
\usepackage{ulem}
\usepackage[dvipsnames]{xcolor}
\renewcommand{\sout}{\bgroup \color{red} \ULdepth=-.5ex \ULset}

\makeatother

\begin{document}

\preprint{INHA-NTG-09/2018}

\title{New narrow nucleon resonances $N^{\ast}(1685)$ and
  $N^{\ast}(1726)$ within the chiral quark-soliton model}

\author{Ghil-Seok Yang}
\email[E-mail: ]{ghsyang@ssu.ac.kr}
\affiliation{Department of Physics, Soongsil University, Seoul 06978,
  Republic of Korea}

\author{Hyun-Chul Kim} 
\email[E-mail: ]{hchkim@inha.ac.kr}
\affiliation{Department of Physics, Inha University, Incheon 22212,
  Republic of Korea} 
\affiliation{Advanced Science Research Center, Japan Atomic Energy 
  Agency, Shirakata, Tokai, Ibaraki, 319-1195, Japan}
\affiliation{School of Physics, Korea Institute for Advanced Study
  (KIAS), Seoul 02455, Republic of Korea}

\date{\today} 
\begin{abstract}
We investigate the strong and radiative decay widths of
the narrow nucleon resonances $N^*(1685)$ and
$N^{\ast}(1726)$ within the framework of the SU(3) chiral
quark-soliton model.  All the relevant parameters are taken from those 
used to describe the properties of the baryon octet and decuplet in
previous works. The masses of the antidecuplet nucleon and
the eikosiheptaplet (27-plet) nucleon with spin 3/2 are determined
respectively to be $(1690.2\pm 10.5)\, \mathrm{MeV}$ and
$(1719.6\pm7.4)\,\mathrm{MeV}$.  The decay width for
$N^{\ast}(1685)\to \eta + N$ is found to be
approximately 
three times larger than that for $N^{\ast}(1685)\to \pi +
N$. The width of the decay $N^{\ast}\left(1726\right)3/2^+\to \eta +
N$ is even about 31 times larger than that of 
$N^{\ast}\left(1726\right)3/2^+\to \pi + N$. The ratio of the
radiative decays for $N^*(1685)$ is obtained to be 
$\Gamma_{nn^*(1685)}/\Gamma_{pp^*(1685)}=8.62\pm3.45$ which explains very
well the neutron anomaly. In contrast, we find
$\Gamma_{pp^*(1726)}/\Gamma_{nn^*(1726)}=3.72\pm0.64$, which indicates
that the production of $N^*(1726)$ is more likely to be observed in the
proton channel. We also examined the decay modes of these narrow
nucleon resonances with the strangeness hadrons involved. 
\end{abstract}



\maketitle
\section{Introduction}
Since Kuznetsov et al. announced the finding of a narrow nucleon
resonance or a narrow bump-like structure around the center of mass
energy $W\sim 1.68\,\mathrm{GeV}$ in photoproduction of $\eta$ mesons
off the quasi-free neutron~\cite{Kuznetsov:2006kt}, several
experimental collaborations have subsequently confirmed its
existence~\cite{Miyahara:2007zz, Jaegle:2008ux, Jaegle:2011sw,
  Werthmuller:2013rba, Witthauer:2013tkm, Werthmuller:2014thb}. The
broad nucleon resonance $N(1535,\,1/2^-)$ appears as the dominant one in
$\eta$ photoproduction off the nucleon, so that this narrow state was
placed on the shoulder of the $N(1535,\,1/2^-)$ in the vicinity of
$W\sim 1.68$ GeV. On the other hand, it was experimentally shown that
such a narrow structure does not exist for the
proton~\cite{Bartalini:2007fg, Jaegle:2008ux} or seems to have a
dip-like structure for the proton~\cite{McNicoll:2010qk,
  Jaegle:2011sw}. Recently, A2 Collaboration has measured the double 
polarization observable and the helicity-dependent cross sections for
$\eta$ photoproduction from the quasi-free protons and
neutrons~\cite{Witthauer:2017get}, using a circularly polarized
photon beam. The narrow structure was observed only in the spin-$1/2$
helicity-dependent cross section, which indicates that a spin-1/2
amplitude is most likely related to this narrow
structure. Reference~\cite{Witthauer:2017get} concluded that this 
structure is unambiguously related to the helicity-1/2
amplitude. Moreover, the experimental data being compared with
different model predictions of the angular dependence, a narrow
structure is favored to be interpreted as a narrow $P_{11}$ nucleon
resonance. The pronounced bump-like structure found only for the
neutron is often called \textit{neutron
  anomaly}~\cite{Kuznetsov:2008hj}.

As the narrow bump-like structure was undisputably established by
several experiments, there has been various theoretical
interpretations of it. Shklyar et al.~\cite{Shklyar:2006xw} explained
that it comes from the coupled-channel effects due to $N(1650)1/2^-$
and $N(1710)1/2^+$ based on the unitary coupled-channels effective
Lagrangian approach, while Shyam and Scholten~\cite{Shyam:2008fr}
described it as the interference effects of $N(1535)1/2^-$,
$N(1650)1/2^-$, $N(1710)1/2^+$, and $N(1720)3/2^+$ resonance
contributions, using a coupled-channels $K$-matrix 
method.  Anisovich et al.~\cite{Anisovich:2008wd, Anisovich:2015tla}
interpreted that the narrow bump-like structure arises from the
interference between the $N(1535)1/2^-$ and the $N(1650)1/2^-$. 
On the other hand, D\"oring and Nakayama~\cite{Doring:2009qr}
studied the ratio of the cross section $\sigma_n/\sigma_p$ with the
intermediate meson-baryon states with strangeness and interpreted the
pronounced narrow structure as the effects coming from the opening of
the strangeness channel in intermediate states.  
On the contrary, Refs.~\cite{Choi:2005ki,Choi:2007gy, Suh:2018}
explained the $\gamma n\to \eta n$ reaction very well within an
Effective Lagrangian approach, regarding the narrow structure as the
narrow nucleon resonance $N^*(1685)1/2^+$. 
In particular, Suh et al.~\cite{Suh:2018} was able to describe the
precise experimental data on the helicity-dependent cross sections
from the A2 Collaboration~\cite{Witthauer:2017get}. 
Kuznetsov et al.~\cite{Kuznetsov:2008hj}
analyzed the GRAAL data and found also the evidence of a narrow
structure in the $\Sigma$ beam asymmetry for the 
reaction $\gamma p \to \eta p$ and argued that the evidence of the
narrow structure in the proton channel cannot be explained by those 
coupled-channel effects. In a more recent
work~\cite{Kuznetsov:2017ayk}, Kuznetsov et al. refuted that
interpretation of the pronounced narrow bump-like structure as an
interference effect between the $S$-wave nucleon resonances. The key
point of Ref.~\cite{Kuznetsov:2017ayk} is that the signal of $N^*(1685)$
may appear in polarization observables for $\eta$ photoproduction off
the proton even if this resonance has a weaker transition magnetic
moments than that of the neutron. We also want to mention that both
the narrow excited proton and neutron were also seen in Compton
scattering $\gamma N\to \gamma N$~\cite{Kuznetsov:2010as,
  Kuznetsov:2015nla} and the reactions $\gamma N\to \pi\eta
N$~\cite{Kuznetsov:2017xgu}. 

In the meanwhile, the second narrow peak around $W\sim 1.72$ GeV was
seen in several different experiments. The beam asymmetry
for $\eta$ photoproduction off the proton~\cite{Witthauer:2017get}
gave a hint for the existence of the second narrow resonance. That for
Compton scattering on the proton exhibited also it together with the first
bump-like structure~\cite{Kuznetsov:2015nla}. Gridnev et
al.~\cite{Gridnev:2016dba} performed the high-precision analysis of
the $\pi^\pm p$ cross-sectional data from the EPECUR Collaboration
based on the multi-channel $K$-matrix approach and found both the
narrow structures at $W\sim 1.68$ GeV and $W\sim 1.72$ GeV,
respectively. What is interesting is that as pointed out by
Werthm\"uller et al.~\cite{Werthmuller:2015owc} the second narrow peak
is much weaker in the $\gamma n \to \eta n$ reaction than $\eta$
photoproduction off the proton. So, the situation is quite opposite to
the case of the first narrow bump-like structure around $W\sim 1.68$
GeV, that is to say there exists an \textit{proton anomaly} in the
case of the second narrow peak.

Mart proposed a possible finding of the narrow nucleon resonance in
$K^0\Lambda$ photoproduction off the neutron~\cite{Mart:2011ey,
  Mart:2013fia, Mart:2017xtf}, though the suggested value of the mass
is lower than 1.68 GeV. Very recently, the FOREST 
Collaboration~\cite{Tsuchikawa:2016ixc, Tsuchikawa:2017tqm}  
carried out the measurement of $K^0\Lambda$ photoproduction off the
quasi-free neutron and the CLAS Collaboration~\cite{Compton:2017xkt,
  Ho:2018riy} have reported the data on the total and 
differential cross sections of $\gamma d \to K^0 \Lambda (p)$. Kim et
al.~\cite{Kim:2018qfu} were able to explain the 
experimental data on both the total and differential cross sections
with the narrow nucleon resonance $N(1685)1/2^+$ included. The
results of the differential cross section revealed the effects of the
$N^*(1685)1/2^+$ in the forward direction near threshold. 

Polyakov and Rathke~\cite{Polyakov:2003dx} demonstrated based on the 
chiral quark-soliton model ($\chi$QSM)~\cite{Diakonov:1987ty,
  Christov:1995vm, Diakonov:1997sj} that the dipole magnetic
transitions between the antidecuplet ($[\overline{\bm{10}}]$) nucleon
and the octet ($[\bm{8}]$) nucleon has a large isospin asymmetry,
i.e. the ratio becomes
$\mu_{nn_{\overline{\bm{10}}}}/\mu_{pp_{\overline{\bm{10}}}}>2$ at
least. 
 Thus, if one assumes that the narrow resonance $N^*(1685)$ belongs to
 the baryon 
antidecuplet, the large isospin asymmetry of the magnetic dipole
transitions can bring about the neutron anomaly in $\eta$
photoproduction off the nucleons. A more quantitative studies on the
dipole magnetic transitions were carried out in
Refs.~\cite{Kim:2005gz, Kim:2007hu, Yang:2013tka} within the same
framework but with the SU(3) symmetry breaking effects
considered. Using also the same framework, Prasza{\l}owicz and 
Goeke~\cite{Praszalowicz:2007zza} examined the masses and strong 
decay widths of the baryon eikosiheptaplet ($[\bm{27}]$). If one assumes
that the second narrow peak belongs to the baryon eikosiheptaplet, the
prediction of its mass is in qualitative agreement with the data. 
However, certain uncertainties were unavoidable in these mentioned 
previous works, so the results could not be determined unambiguously.  
In Refs.~\cite{Yang:2010fm, Yang:2011qe}, the masses of the
lowest-lying baryons including the decuplet and
antidecuplet were unequivocally determined by including the effects of
isospin symmetry breaking arising from both the difference of the
quark masses and the electromagnetic (EM)
self-energies~\cite{Yang:2010id}. These works allowed one to determine 
all the relevant parameters for the vector and axial-vector
properties of the baryons~\cite{Yang:2015era} even including those of
heavy baryons~\cite{Kim:2017khv, Yang:2018uoj}.  
Thus, in the present work, we investigate both the strong and
radiative decay widths of both the pronounced narrow resonances within 
the $\chi$QSM, assuming that the first narrow bump-like structure or
$N^*(1685)1/2^+$ belongs to the baryon antidecuplet and the second
narrow peak or $N^*(1726)3/2^+$ to the baryon eikosiheptaplet. 
The great virtue of the present work is that we do not have any
additional parameters to adjust, since all relevant dynamical
parameters have been fixed in Refs.~\cite{Yang:2010id, Yang:2010fm,
  Yang:2011qe, Yang:2015era}.  
We will demonstrate the following four significant points in this
work, all of which were relevant to the existing experimental
observations: 
\begin{itemize}
\item The second narrow peak can most likely be identified as
  $N^*(1726)3/2^+$. The masses of those with spin 1/2 turn out be larger
  than 2 GeV. The predicted mass of $N^*(1726)3/2^+$ is
  $M_{N_{\bm{27}}}=(1719.6\pm7.4)$ MeV.  
\item Both the narrow nucleon resonances $N^*(1685)1/2^+$ and
  $N^*(1726)3/2^+$ can be found more clearly in the $\eta$ channel than
  the pion channel, since the values of the strong decay widths
  $\Gamma_{\eta N}$ are respectively 2.6 and 31 times larger than
  those of $\Gamma_{\pi N}$. 
\item The ratio of the radiative decay widths for $N^*(1685)1/2^+$ turns
  out to be
  $\Gamma_{n_{\overline{\bm{10}}}(1685)\,n}/\Gamma_{p_{\overline{\bm{10}}}(1685)\,p}
  = 8.62$,  
  which  explains the reason for the neutron anomaly.
The ratio of the radiative decay widths for $N^*(1726)3/2^+$ turns
  out to be
  $\Gamma_{p_{\bm{27}}(1726)\,p}/\Gamma_{n_{\bm{27}}(1726)\,n} =
  3.76$, which explains the reason for the proton anomaly, though the
  result is not as prominent as that of $N^*(1685)1/2^+$.
\item The $N_{\bm{27}}$ is more likely to be observed in decays with
  strangeness than $N_{\overline{\bm{10}}}$.
\end{itemize}

The present paper is organized as follows: In Section II, we briefly
recapitulate the general formalism of the $\chi$QSM, focusing on the
strong and radiative decays of $N^*(1685)$ and $N^*(1726)$. In Section
III, we present the predicted masses of the antidecuplet and
eikosiheptaplet nucleons. In Section IV, we show the results of the
strong and radiative decays of $N^*(1685)$ and $N^*(1726)$ and discuss
them in the context of recent experimental data. 
In the final Section, we summarize the present work and draw
conclusions.  
\section{Collective hamiltonian and wavefunctions of 
SU(3) baryons \label{sec:genformalism}}
\begin{figure}[htp]
\caption{Weight diagrams for the lowest-lying baryon multiplets: from
  the left, the baryon octet, decuplet, antidecuplet and
  eikosiheptaplet (${\bm{27}}$-plet).}\label{fig:1}  
\includegraphics[scale=0.75]{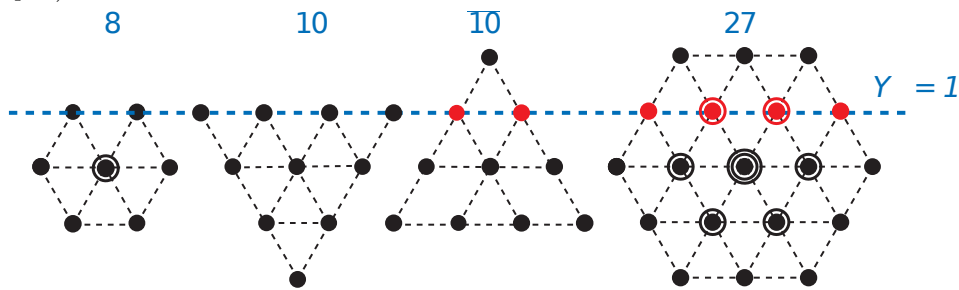} 
\end{figure}

In the $\chi$QSM, the SU(3) soliton is built from a hedgehog ansatz
which requires the embedding of the SU(2) soliton into
SU(3)~\cite{Witten:1983tw}. Then the semiclassical quantization can be
performed by rotating the soliton in both configuration space and
flavor space simultaneously. Since the SU(2) soliton commutes with the
hypercharge operator, we have only the seven zero rotational
modes. Thus, the hypercharge operator does not correspond to any zero
mode. It means that eight component of the generalized momenta
conjugate to the right angular velocities turns out to be constrained,
so that the right hypercharge becomes $Y'= N_cB/3=1$, where $N_c$ is
the number of colors and $B$ is the baryon number. The constraint on
the right hypercharge is imposed by the valence quark in the 
$\chi$QSM~\cite{Christov:1995vm, Diakonov:1997sj}, whereas 
it comes from the Wess-Zumino term in the Skyrme
model~\cite{Chemtob:1985ar, Mazur:1984yf, Jain:1984gp}. The right
hypercharge $Y'=1$ allows the baryon representations that contain
those with right hypercharge $Y'=1$. If the number of the baryons with
$Y'=1$ is expressed as $2J+1$, then the spin of the allowed multiplet
should be equal to $J$ (see Fig.~\ref{fig:1}).  

Having performed the SU(3) zero-mode quantization, we can express the
collective Hamiltonian as
\begin{align}
H & =  
M_{\mathrm{cl}}\;+\;H_{\mathrm{rot}}\;+\;H_{\mathrm{sb}}^{\mathrm{iso}}
\;+\;H_{\mathrm{sb}}^{\mathrm{SU(3)}},
\label{eq:ham}
\end{align}
where $M_{\mathrm{cl}}$ denotes the classical soliton mass. The
$1/N_c$ rotational term, the isospin breaking one, and the SU(3)
symmetry breaking one are respectively written as 
\begin{align}
H_{\mathrm{rot}} & = \frac{1}{2I_{1}}\sum_{i=1}^{3}\hat{J}_{i}^{2} 
\;+\;\frac{1}{2I_{2}}\sum_{p=4}^{7}\hat{J}_{p}^{2},
\label{eq:hamrot}  \\
H_{\mathrm{sb}}^{\mathrm{iso}} 
& =  
\Delta_{\mathrm{du}} \left(\frac{\sqrt{3}}{2}
\,\alpha\, D_{38}^{(\bm{8})}(A)\;+\;\beta\,\hat{T_{3}}
\;+\;\frac{1}{2}\,\gamma\sum_{i=1}^{3}D_{3i}^{(\bm{8})}(A)\,\hat{J}_{i}\right),
\cr
H_{\mathrm{sb}}^{\mathrm{SU(3)}} 
& =  
\Delta_{\mathrm{s}}\left(\alpha\, D_{88}^{(\bm{8})}(A) 
\;+\;\beta\,\hat{Y}\;+\;\frac{1}{\sqrt{3}}
\,\gamma\sum_{i=1}^{3}D_{8i}^{(\bm{8})}(A)\,\hat{J}_{i}\right)
+(m_{\mathrm{u}} + m_{\mathrm{d}}  + m_{\mathrm{s}} )\sigma,
\label{eq:hbr}
\end{align}
where $I_1$ and $I_2$ stand for the moments of
inertia of the soliton~\cite{Blotz:1992pw, Christov:1995vm}. $J_i$ and
$J_p$ denote the generators of the SU(3) group. $\Delta_{\mathrm{du}}$
and $\Delta_{\mathrm{s}}$ are defined by $\Delta_{\mathrm{du}} =
m_{\mathrm{d}} - m_{\mathrm{u}}$ and $\Delta_{\mathrm{s}} =
m_{\mathrm{s}} - \overline{m}$, where $m_{\mathrm{u}}$,
$m_{\mathrm{d}}$, and $m_{\mathrm{s}}$ designate the current quark
masses with the corresponding flavor. $\overline{m}$ is the average
current mass of the up and down quarks,
$\overline{m}=(m_{\mathrm{u}}+m_{\mathrm{d}})/2$.
We are not able to determine separately the model parameters $\alpha$,
$\beta$, and $\gamma$ but we do not need to separate them, since we
can determine all these parameters by using the masses of the baryon
octet, the $\Omega^-$ mass, and that of the putative $\Theta^+$. 
$\sigma$ is defined as 
\begin{align}
  \label{eq:sigma}
\sigma = -(\alpha+\beta) = \frac13 \frac{\Sigma_{\pi N}}{\overline{m}},
\end{align}
where $\Sigma_{\pi N}$ stands for the $\pi N$ sigma term. 
The parameters in Eq.~\eqref{eq:hbr} have been already determined in
Ref.~\cite{Yang:2010fm}:  
\begin{align}
\Delta_{\mathrm{du}} \alpha 
& =  -4.390\pm0.004,\;\;\;\;\;
\Delta_{\mathrm{s}}\alpha\;=\;-255.029\pm5.821,
\cr
\Delta_{\mathrm{du}}\beta 
& =  -2.411\pm0.001,\;\;\;\;\;\;\;
\Delta_{\mathrm{s}}\beta\;=\;-140.040\pm3.195,
\cr
\Delta_{\mathrm{du}}\gamma 
& =  -1.740\pm0.006,\;\;\;\;\;\;\;
\Delta_{\mathrm{s}}\gamma\;=\;-101.081\pm2.332,
\label{eq:Nabrfinal}
\end{align}
in units of MeV. We will use these values to calculate the masses of
the $N^*(1685)$ and $N^*(1726)$. $D_{ij}^{(8)}(A)$ in Eq.~\eqref{eq:hbr}
denotes the SU(3) Wigner $D$ functions. 

In the  $(p,\,q)$ representation of the SU(3) group, the sum of the
generators can be expressed in terms of $p$ and $q$
\begin{align}
\sum_{i=1}^{8}J_{i}^{2}=\frac{1}{3}\left[p^{2}+q^{2}+p\;q+3(p+q)\right],
\label{eq:Jsquare}
\end{align}
which yields the eigenvalues of the rotational collective Hamiltonian
$H_{\mathrm{rot}}$ in Eq.~\eqref{eq:hamrot} as follows: 
\begin{align}
E_{(p,\,q),\,J} & =  
\mathcal{M}_{\mathrm{cl}}
\;+\;\frac{1}{2}
\left(\frac{1}{I_{1}}\,-\,\frac{1}{I_{2}}\right)\,J\,(J\,+\,1)
 +\frac{1}{6I_{2}}
\left(p^{2}\,+\,q^{2}\,+\,3(p\,+\,q)\,+\,p\,q\right)\,-\,\frac{3}{8I_{2}}.
\label{eq:EpqJ}
\end{align}
Then the allowed SU(3) baryon multiplets with zero triality are
obtained as 
\begin{align}
(p,\,q) & =  (1,\,1)\;\;
\rightarrow\;\;J=1/2\;\;(\mbox{octet}),
\cr
(p,\,q) & =  (3,\,0)\;\;
\rightarrow\;\;J=3/2\;\;(\mbox{decuplet}),
\cr
(p,\,q) & =  (0,\,3)\;\;
\rightarrow\;\;J=1/2\;\;(\mbox{antidecuplet}),
\cr
(p,\,q) & = 
(2,\,2)\;\;
\rightarrow\;\;J=1/2,\;J=3/2\;\;(\mbox{eikosiheptaplet}).
\label{eq:pq}
\end{align}
Note that the eikosiheptaplet has two degenerate representations with
$J=1/2$ and $J=3/2$. The mass splittings between the centers of the
multiplets $\overline{M}_{\mathbf{\mathcal{R}}}^{J},\;
\left(E_{(p,q)}^{J} \right)$ are determined by the moments of inertia  
\begin{align}
\overline{M}_{\mathbf{\overline{10}}}-\overline{M}_{\mathbf{8}}\;\,
&= 
 \;E_{(0,3)}^{1/2}\, - \,E_{(1,1)}^{1/2}  =\;\;\frac{3}{2\,I_{2}},
\cr
\overline{M}_{\mathbf{27}}^{1/2} - \overline{M}_{\mathbf{8}}\;\,
&= 
 E_{(2,2)}^{1/2}\, - \,E_{(1,1)}^{1/2}  =\;\;\frac{5}{2\,I_{2}},
\cr
\overline{M}_{\mathbf{27}}^{3/2}-\overline{M}_{\mathbf{8}}\;\,
&= 
 E_{(2,2)}^{3/2}\, - \,E_{(1,1)}^{1/2} =\;\;\frac{3}{2\,I_{1}}+\frac{1}{I_{2}},
\label{eq:centerdiff}
\end{align}
where the center mass values of the antidecuplet and eikosiheptaplet
are obtained as 
\begin{align}
\overline{M}_{ \overline{\bm{10}}} \;\;\;\;
 =  1854.9\pm10.0\;\mathrm{MeV},
\;\;\;
\overline{M}_{\bm{27}}^{1/2} 
 =  2324.7\pm16.7\;\mathrm{MeV},
\;\;\;
\overline{M}_{\bm{27}}^{3/2} 
 =  1860.4\pm6.7\;\mathrm{MeV}.
\label{eq:centervalue}
\end{align}

The baryon wavefunctions for ta representation ${\bm{\mathcal{R}}}$ are
written in terms of the SU(3) Wigner $D$
functions~\cite{Yang:2010fm,Blotz:1992pw}:   
\begin{align}
\langle A|{\bm{\mathcal{R}}},\,B(Y\,T\,T_{3},\;Y^{\prime}\,J\,J_{3})\rangle 
=\Psi_{({\bm{\mathcal{R}}}^{*}\,;\,Y^{\prime}\,J\,J_{3})}^{({\bm{\mathcal{R}}}\,;\,Y\,T\,T_{3})}(A)
=\sqrt{\textrm{dim}({\bm{\mathcal{R}}})}\,(-)^{J_{3}+Y^{\prime}/2}\,
D_{(Y,\,T,\,T_{3})(-Y^{\prime},\,J,\,-J_{3})}^{({\bm{\mathcal{R}}})*}(A),
\label{eq:5}
\end{align}
where ${\bm{\mathcal{R}}}$ denotes one of the allowed irreducible
representations in SU(3), i.e. ${\bm{\mathcal{R}}}\,=\,\bm{8},\,\bm{10},\, 
\overline{\bm{10}},\,\bm{27},\cdots$.  $Y,\,T,\,T_{3}$ are the
corresponding hypercharge, isospin and its third component,
respectively. As mentioned previously, the right hypercharge is
constrained to be $Y^{\prime}=1$.

The presence of the SU(3) symmetry breaking term in Eq.~\eqref{eq:hbr}  
will drive a baryon state to be mixed with those from higher
representations. Thus, the wavefunctions for the baryon octet, the
decuplet, the antidecuplet, and the eikosiheptaplet are expressed
respectively as 
\begin{align}
\left|B_{\bm{8}}\right\rangle  
& 
=\left|{\bm{8}}_{1/2},\,B\right\rangle 
\;+\;c_{\overline{\bm{10}}}^{B}
\left|{\overline{\bm{10}}}_{1/2},\,B\right\rangle 
\;+\;c_{\bm{27}}^{B}
\left|{\bm{27}}_{1/2},\,B\right\rangle ,
\cr
\left|B_{\bm{10}}\right\rangle  
& 
=\left|{\bm{10}}_{3/2},\,B\right\rangle
\;+\;a_{\bm{27}}^{B}
\left|{\bm{27}}_{3/2},\,B\right\rangle 
\;+\;a_{\bm{35}}^{B}
\left|{\bm{35}}_{3/2},\,B\right\rangle ,
\cr
\left|B_{\overline{\bm{10}}}\right\rangle  
&
=\left|{\overline{\bm{10}}}{}_{1/2},\,B\right\rangle 
\;+\;d_{{\bm{8}}}^{B}\left|{\bm{8}}_{1/2},\,B\right\rangle
\;+\;d_{\bm{27}}^{B}\left|{\bm{27}}_{1/2},\,B\right\rangle 
\;+\;d_{\overline{\bm{35}}}^{B}\left|{{\overline{\bm{35}}}}_{1/2},\,B\right\rangle,
\cr
\left|B_{\bm{27}}^{1/2}\right\rangle  
& =  
\left|{\bm{27}}_{1/2},\,B\right\rangle
\;+\;n_{\bm{8}}^{B}\left|{\bm{8}}_{1/2},\,B,\right\rangle 
\;+\;n_{\overline{\bm{10}}}^{B}\left|{\overline{\bm{10}}}_{1/2},\,B\right\rangle 
\;+\;n_{\overline{\bm{35}}}^{B}\left|{\overline{\bm{35}}}_{1/2},\,B\right\rangle 
\;+\;n_{\bm{64}}^{B}\left|{\bm{64}}_{1/2},\,B\right\rangle ,
\cr
\left|B_{{\bm{27}}}^{3/2} \right\rangle  & =  
\left|{\bm{27}}_{3/2},\,B\right\rangle 
\;+\;m_{\bm{10}}^{B}\left|{\bm{10}}_{3/2},\,B\right\rangle 
\;+\;m_{\bm{35}}^{B}\left|{\bm{35}}_{3/2},\,B\right\rangle 
\;+\;m_{\overline{\bm{35}}}^{B}\left|{\overline{\bm{35}}}_{3/2},\,B\right\rangle 
\;+\;m_{\bm{64}}^{B}\left|{\bm{64}}_{3/2},\,B\right\rangle,
\label{eq:su3wf}
\end{align}
where the mixing coefficients in Eq.~\eqref{eq:su3wf} are presented in
Appendix~\ref{sec:Appendix1}. Note that
$\left|B_{{\bm{\mathcal{R}}}} ^J\right\rangle $ will be reduced 
to the pure state $\left|{\bm{\mathcal{R}}}_J,\,B\right\rangle $ of
Eq.~\eqref{eq:5}, if one takes the limit of
$m_{\mathrm{s}}\rightarrow0$. 

The results of the mass splittings for the baryon decuplet and
antidecuplet have been already presented and discussed in
Ref.~\cite{Yang:2010fm}. However, we recapitulate them in the present
work in Tables~\ref{tab:Octm1}, \ref{tab:decm1}, and \ref{tab:antim1}
given below. We use the masses of the $\Omega^-$ and $\Theta^+$ as 
input together with those of the baryon octet. We could use the
mass of the $N_{\overline{10}}$ as input instead of $\Theta^+$ but
it is more reasonable to choose that of $\Theta^+$ since it is the
isosinglet. By the same token, we choose the mass of $\Omega^-$ as
input to determine the masses of all other members of the baryon
decuplet. In addition, we also considered the contributions of the
isospin symmetry breaking such that we are able to use all the masses
of the baryon octet as input. There are two different contributions on
the isospin symmetry breaking: the hadronic part and the
electromagnetic part, both of which were included in the calculation
of the mass splittings of the lowest-lying SU(3) baryons. We find that
the present scheme yields robust results. In particular, we are able
to reproduce the masses of the baryon decuplet without any additional
parameters introduced. 

\begin{table}[th]
 \centering \caption{Reproduced masses of the baryon octet. The
   experimental data of octet baryons are taken from the Particle 
Data Group (PDG).} 
\begin{tabular}{cccccc}
\hline 
\multicolumn{2}{c}{Mass {[}MeV{]}} & $T_{3}$  & $Y$  
& Exp. {[}Inputs{]}  & Numerical results
\tabularnewline
\hline 
$M_{N}$  & $\begin{array}{c}
p\\
n
\end{array}$  & $\begin{array}{c}
\;\;1/2\\
-1/2
\end{array}$  & $\;\;1$  & $\begin{array}{c}
938.27203\pm0.00008\\
939.56536\pm0.00008
\end{array}$  & $\begin{array}{c}
938.76\pm3.65\\
940.27\pm3.64
\end{array}$\tabularnewline
\hline 
$M_{\Lambda}$  & $\Lambda$  & $\;\;0$  
& $\;\;0$  & $1115.683\pm0.006$  & $1109.61\pm0.70$
\tabularnewline
\hline 
$M_{\Sigma}$  & $\begin{array}{c}
\Sigma^{+}\\
\Sigma^{0}\\
\Sigma^{-}
\end{array}$  & $\begin{array}{c}
\;\;1\\
\;\;0\\
-1
\end{array}$  & $\;\;0$  & $\begin{array}{c}
1189.37\;\pm0.07\\
1192.642\pm0.024\\
1197.449\pm0.030
\end{array}$  & $\begin{array}{c}
1188.75\pm0.70\\
1190.20\pm0.77\\
1195.48\pm0.71
\end{array}$\tabularnewline
\hline 
$M_{\Xi}$  & $\begin{array}{c}
\Xi^{0}\\
\Xi^{-}
\end{array}$  & $\begin{array}{c}
\;\;1/2\\
-1/2
\end{array}$  & $-1$  & $\begin{array}{c}
1314.83\pm0.20\\
1321.31\pm0.13
\end{array}$  & $\begin{array}{c}
1319.30\pm3.43\\
1324.52\pm3.44
\end{array}$\tabularnewline
\hline 
\end{tabular}\label{tab:Octm1} 
\end{table}

\begin{table}[th]
 \centering \caption{Predicted masses of the baryon decuplet. The
   experimental data of octet baryons are taken from the Particle Data
   Group (PDG).} 
\begin{tabular}{cccccc}
\hline 
\multicolumn{2}{c}{Mass {[}MeV{]}} 
& $T_{3}$  
& $Y$  
& Experiment \cite{Tanabashi:2018oca}  
& Predictions
\tabularnewline
\hline 
$M_{\Delta}$  
& $\begin{array}{c}
\Delta^{++}\\
\Delta^{+}\\
\Delta^{0}\\
\Delta^{-}
\end{array}$  
& $\begin{array}{c}
\;\;3/2\\
\;\;1/2\\
-1/2\\
-3/2
\end{array}$  
& $\;\;1$  
& $1231-1233$  
& $\begin{array}{c}
1248.54\pm3.39\\
1249.36\pm3.37\\
1251.53\pm3.38\\
1255.08\pm3.37
\end{array}$\tabularnewline
\hline 
$M_{\Sigma^{\ast}}$  
& $\begin{array}{c}
\Sigma^{\ast+}\\
\Sigma^{\ast0}\\
\Sigma^{\ast-}
\end{array}$  
& $\begin{array}{c}
\;\;1\\
\;\;0\\
-1
\end{array}$  
& $\;\;0$  
& $\begin{array}{c}
1382.8\pm0.4\;\\
1383.7\pm1.0\\
1387.2\pm0.5\,
\end{array}$  
& $\begin{array}{c}
1388.48\pm0.34\\
1390.66\pm0.37\\
1394.20\pm0.34
\end{array}$
\tabularnewline
\hline 
$M_{\Xi^{\ast0}}$  
& $\begin{array}{c}
\Xi^{\ast0}\\
\Xi^{\ast-}
\end{array}$  
& $\begin{array}{c}
\;\;1/2\\
-1/2
\end{array}$  
& $-1$  
& $\begin{array}{c}
1531.80\pm0.32\\
1535.0\;\pm0.6\;\,
\end{array}$  
& $\begin{array}{c}
1529.78\pm3.38\\
1533.33\pm3.37
\end{array}$
\tabularnewline
\hline 
$M_{\Omega^{-}}^{\star}$  
& $\Omega^{-}$  
& $0$  & $-2$  
& $1672.45\pm0.29$  
& Input
\tabularnewline
\hline 
\end{tabular}\label{tab:decm1} 
\end{table}

\begin{table}[h]
 \centering \caption{Predicted masses of the baryon antidecuplet.}
\begin{tabular}{cccccc}
\hline 
\multicolumn{2}{c}{Mass } 
& $T_{3}$  
& $Y$  
& Experiment  
& Predictions
\tabularnewline
\hline 
$M_{\Theta^{+}}$  
& $\Theta^{+}$  
& $\;\;0$  
& $\;\;2$  
& $1524\pm 5$\cite{Nakano:2008ee}  
& Input
\tabularnewline
\hline 
$M_{N_{\overline{10}}}$  
& $\begin{array}{c}
p_{\overline{10}}\\
n_{\overline{10}}
\end{array}$  
& $\begin{array}{c}
\;\;1/2\\
-1/2
\end{array}$  
& $\;\;1$  
& ${\displaystyle 1686\pm 12}$
\cite{Kuznetsov:2008hj}
& $\begin{array}{c}
1688.18\pm10.53\\
1692.16\pm10.53
\end{array}$\tabularnewline
\hline 
$M_{\Sigma_{\overline{10}}}$  
& $\begin{array}{c}
\Sigma_{\overline{10}}^{+}\\
\Sigma_{\overline{10}}^{0}\\
\Sigma_{\overline{10}}^{-}
\end{array}$  
& $\begin{array}{c}
\;\;1\\
\;\;0\\
-1
\end{array}$  
& $\;\;0$  
&  
& $\begin{array}{c}
1852.35\pm10.00\\
1856.33\pm10.00\\
1858.95\pm10.00
\end{array}$\tabularnewline
\hline 
$M_{\Xi_{3/2}}$  
& $\begin{array}{c}
\Xi_{3/2}^{+}\\
\Xi_{3/2}^{0}\\
\Xi_{3/2}^{-}\\
\Xi_{3/2}^{--}
\end{array}$  
& $\begin{array}{c}
\;\;3/2\\
\;\;1/2\\
-1/2\\
-3/2
\end{array}$  
& $\;\;-1$  
&  
& $\begin{array}{c}
2016.53\pm10.53\\
2020.51\pm10.53\\
2023.12\pm10.53\\
2024.37\pm10.53
\end{array}$\tabularnewline
\hline 
\end{tabular}
\label{tab:antim1} 
\end{table}
In Table~\ref{tab:Octm1}, we list the reproduced masses of
the baryon octet. In Table~\ref{tab:decm1}, we present the results of the
decuplet masses. We find that the masses of the $\Sigma^{\ast}$ and
$\Xi^{\ast}$ are remarkably in good agreement with the data within
$0.5\,\%$. Table~\ref{tab:antim1} lists the predictions of the
antidecuplet masses. In fact, the existence of $\Theta^+$ is
putative and controversial. Even though it exists, it is still very
difficult to measure it experimentally, because of its small decay
width. For example, the DIANA Collaboration announced recently the 
decay width of $\Theta^+$ to be $0.34\pm 0.10$
MeV~\cite{Barmin:2013lva}. Though the existence of the baryon
antidecuplet is questionable because of lack of experimental data, we
hope that future experiments at the J-PARC with the kaon beam may put
a period on the matter whether $\Theta^+$ exists or
not~\cite{Sekihara}. So far, the masses of the
$\Sigma_{\overline{10}}$ were not observed experimentally. One can
explain a possible reason as follows: Since the
$\Sigma_{\overline{10}}$ has the same strangeness as usual 
hyperons, it is rather difficult to distinguish them from other
excited hyperons within the range from $1.8$ GeV to $2.0$ GeV. When it
comes to the $\Xi_{3/2}$, one of the corresponding multiplet has
double negative charge. In fact, The finding of $\Xi_{3/2}$ was
reported by the NA49 Collaboration~\cite{Alt:2003vb}, though it was
not confirmed by other experiments. Moreover, the $\Xi_{3/2}$ mass
observed by the NA49 Collaboration, $1.862\pm0.002$ GeV, is quite
different from the predictions of the present work as shown in
Table~\ref{tab:antim1}. 

On the other hand, $N_{\overline{10}}(1685)$ is experimentally found
to decay exclusively into $\eta N$, which is a unique feature in
comparison with all other $N^*$ resonances except for the
$N^*(1535)$. Thus, assuming that $N_{\overline{10}}(1685)$ belongs to
the baryon antidecuplet, we can systematically investigate its
properties within the present framework. In the present work, we will
focus on the $N_{\overline{10}}$ and $N_{27}$, regarding them as the
pentaquark baryons. We will keep all the parameters the same as in the
previous works. In the next Section, we first discuss the results of
the $N_{\overline{10}}$ and $N_{27}$ masses and then we continue to 
examine the radiative and strong decays of them. 

\section{Masses of the antidecuplet and 
eikosiheptaplet nucleons} 
The centers of mass splittings $\overline{M}_{\bm{27}}^{J}$ can  be
easily determined by the center mass of the octet baryon and the
soliton moments of inertia $I_{1,\,2}$, since the splittings between
different representations are provided with the rotational
excitations:  
\begin{eqnarray}
\overline{M}_{\bm{27}}^{1/2} 
& = & 
\overline{M}_{\bm{8}}\;+\;\frac{5}{2I_{2}},
\cr
\overline{M}_{\bm{27}}^{3/2} 
& = & 
\overline{M}_{\bm{8}}\;+\;\frac{3}{2I_{1}}+\frac{1}{I_{2}},
\label{eq:27centers}
\end{eqnarray}
where $\overline{M}_{\bm{8}}$ denotes the center mass of the baryon
octet derived in Ref.\cite{Yang:2010fm}. 
In order to compute the masses of the antidecuplet and eikosiheptaplet
nucleons quantitatively, we need to include the EM self-energies in
addition to the isospin symmetry breaking arising from the difference
between the current up and down quark masses and the SU(3) symmetry
breaking term given in Eq.~\eqref{eq:hbr}. In
Refs.~\cite{Yang:2010id, Yang:2010fm}, the collective operator for
the EM self-energies are given by 
\begin{align}
  \label{eq:EM}
\mathcal{O}^{\mathrm{EM}} = \delta^{(27)} \left(\sqrt{5} 
D_{\Sigma_2^0 \Lambda_{27}}^{(27)} + \sqrt{3}  
D_{\Sigma_1^0  \Lambda_{27}}^{(27)} 
+ D_{\Lambda_{27}  \Lambda_{27}}^{(27)} \right) 
+ \delta^{(8)} \left(\sqrt{3} D_{\Sigma^0  \Lambda}^{(8)} 
+ D_{\Lambda  \Lambda}^{(8)}\right) + \delta^{(1)} D_{\Lambda  \Lambda}^{(1)}, 
\end{align}
where $\delta^{(n)}$ with $n=1,\,8,\,27$ encodes specific dynamics of
the $\chi$QSM. We can determine $\delta^{(27)}$ and $\delta^{(8)}$ by
using the empirical data estimated in
Ref.~\cite{Gasser:1982ap}. $\delta^{(1)}$ 
can be absorbed in the center masses. Sandwiching the operator in
Eq.~\eqref{eq:EM} between the nucleon states, we can get the
corrections of the EM self-energies to the masses of the antidecuplet
and eikosiheptaplet nucleons. The explicit expressions can be found in
Table~\ref{tab:4} in Appendix~\ref{app:b}. The effects of the isospin
symmetry breaking arising from the current quark mass difference can
be also obtained in a similar manner. See also Table~\ref{tab:4} in
Appendix~\ref{app:b} for the expressions of the corrections of the the
isospin symmetry breaking due to the current quark mass difference.   
Concerning the effects of the flavor SU(3) symmetry breaking, we
compute the matrix elements of $H_{\mathrm{sb}}^{\mathrm{SU(3)}}$ in
Eq.~\eqref{eq:hbr}. The explicit expressions for the SU(3) symmetry
breaking part can be found also in Table~\ref{tab:4}. 

\begin{table}[htp]
\global\long\def\arraystretch{1.5}
\caption{Predicted values of the masses for the antidecuplet
and eikosiheptaplet nucleons in units of MeV.}
\label{tab:1} %
\begin{tabular}{c|ccrcc}
\hline
\multicolumn{3}{c}{States} 
& {$T_{3}\;\;$}  
& Mass  
& Average 
\cr
\hline 
\multirow{2}{*}{${\overline{\bm{10}}}_{1/2}$}  
& \multirow{2}{*}{$N_{\overline{\bm{10}}}$}  
& $p_{\overline{\bm{10}}}$  
& {$1/2$}  
& $1688.2\pm10.5$  
& \multirow{2}{*}{$1690.2\pm10.5$} 
\cr
 &  
& $n_{\overline{\bm{10}}}$  
& {$-1/2$}  
& $1692.2\pm10.5$  
& \cr
\hline 
\multicolumn{6}{c}{\vspace{-1.8em}
}\cr
\hline 
\multirow{2}{*}{${\bm{27}}_{1/2}$}  
& \multirow{2}{*}{{$N_{{\bm{27}}}$}}  
& {$p_{{\bm{27}}}$}  
& {$1/2$}  
& $2115.7\pm17.0$  
& \multirow{2}{*}{$2116.6\pm17.0$} 
\cr
 &  
& {$n_{{\bm{27}}}$}  
& {$-1/2$}  
& $2117.4\pm17.0$  
& \cr
\multicolumn{6}{c}{\vspace{-1.8em}
}\cr
\hline 
\multirow{2}{*}{${\bm{27}}_{3/2}$}  
& \multirow{2}{*}{{$N_{{\bm{27}}}$}}  
& {$p_{{\bm{27}}}$}  
& {$1/2$}  
& $1718.6\pm7.4$  
& \multirow{2}{*}{$1719.6\pm7.4$} 
\cr
 &  
& {$n_{{\bm{27}}}$}  
& {$-1/2$}  
& $1720.6\pm7.4$  
& \cr
\hline 
\end{tabular}
\end{table}
In Table~\ref{tab:1}, we list the numerical results of the
antidecuplet and eikosiheptaplet nucleons in units of MeV. 
We want to emphasize again that these values have been produced by
using the parameters given in Eq.~\eqref{eq:Nabrfinal}. 
The effects of isospin symmetry breaking are stronger on $N^*(1685)$
than on $N_{\bm{27}}$'s. In general, the EM effects ($\Delta 
M_{N_{{\bm{\mathcal{R}}}}}^{\mathrm{EM}}$) are approximately two times
smaller than the current quark mass difference ($\Delta
M_{N_{{\bm{\mathcal{R}}}}}^{d-u}$). We get for the 
antidecuplet neutron $\Delta
M_{n_{\overline{\bm{10}}}}^{\mathrm{EM}}=(0.7\pm 0.2)$ MeV and  
$\Delta M_{n_{\overline{\bm{10}}}}^{d-u}=1.4$
MeV, and for $p_{\overline{\bm{10}}}$ we find $\Delta
M_{p_{\overline{\bm{10}}}}^{\mathrm{EM}}=(-0.4\pm 0.2)$ MeV and  
$\Delta M_{p_{\overline{\bm{10}}}}^{d-u}=-1.4$
MeV. On the other hand, we 
obtain $\Delta M_{n_{\bm{27}}}^{\mathrm{EM}} (J=1/2) = (-1.0\pm 0.3)$ MeV and 
$\Delta M_{n_{\bm{27}}}^{d-u} (J=1/2) = 1.2$ MeV, $\Delta
M_{p_{\bm{27}}}^{\mathrm{EM}}(J=1/2) = (-0.5\pm 0.3)$ MeV, and  
$\Delta M_{p_{\bm{27}}}^{d-u} (J=1/2) = -1.2$ MeV. So, the EM effects on
$n_{\bm{27}}(J=1/2)$ is canceled by the corresponding isospin symmetry
breaking effects in the case of the eikosiheptaplet neutron. As a
result, the isospin mass difference between the eikosiheptaplet
neutron and the corresponding proton turns out to be smaller than that
between $n_{\overline{\bm{10}}}$ and $p_{\overline{\bm{10}}}$.
 
The predicted masses of $N^*(1685)$ turn out to 
be slightly larger than those found in experimental data.
Notably, the masses of the $N_{\bm{27}} ({J=3/2})$ are found to be
smaller than those of $N_{\bm{27}}({J=1/2})$, which was already shown
in Ref.~\cite{Praszalowicz:2007zza}, though the values in
Ref.~\cite{Praszalowicz:2007zza} depend on $\Sigma_{\pi N}$. Note that
in the present framework the $\Sigma_{\pi N}$ was also
predicted~\cite{Yang:2010fm}. Thus the second narrow peak observed in 
several experiments can be identified as an eikosiheptaplet nucleon. 
The predicted mass of the eikosiheptaplet neutron is $(1720.6\pm 7.4)$
MeV. From now on, we will denote the notation $N_{\bm{27}}$
exclusively as the eikosiheptaplet nucleon with spin $3/2$ for
compactness.  

\section{Strong and radiative decay widths of 
$N_{\overline{\bm{10}}}$ and $N_{\bm{27}}$ 
\label{sec:results}}
In this Section, we compute the strong and radiative decay widths of
both the antidecuplet and eikosiheptaplet nucleons. 
The collective operators for the axial-vector and magnetic
transitions~\cite{Kim:1997ts, Kim:1997ip, Yang:2015era, Yang:2018idi}  
\begin{align}
  \label{eq:gmu}
  \hat{g}_1 &= \hat{g}_1^{(0)} + \hat{g}_1^{(1)},    \\
\label{eq:mu}
\hat{\mu} &= \hat{\mu}^{(0)} + \hat{\mu}^{(1)}, 
\end{align}
where
\begin{align}
\hat{g}_{1}^{(0)} & 
=\;\;a_{1}\,D_{\varphi3}^{(8)}
\;+\;a_{2}d_{3bc}\,D_{\varphi  b}^{(8)}\,\hat{J}_{c}
\;+\;\frac{a_{3}}{\sqrt{3}}\,D_{\varphi8}^{(8)}\,\hat{J}_{3},
\label{eq:op0} \\
\hat{g}_{1}^{(1)} & =  
\frac{a_{4}}{\sqrt{3}}\,d_{pq3}D_{\varphi p}^{(8)}\,D_{8q}^{(8)}
+a_{5}\,\left(D_{\varphi3}^{(8)}\,D_{88}^{(8)} +
                    D_{\varphi8}^{(8)}\,D_{83}^{(8)}\right) 
+a_{6}\,\left(D_{\varphi3}^{(8)}\,D_{88}^{(8)} -
                    D_{\varphi8}^{(8)}\,D_{83}^{(8)}\right), 
\label{eq:op1} \\
\hat{\mu}^{(0)} & 
=\;\;w_{1}\,D_{Q3}^{(8)}
\;+\;w_{2}d_{3bc}\,D_{Q  b}^{(8)}\,\hat{J}_{c}
\;+\;\frac{w_{3}}{\sqrt{3}}\,D_{Q 8}^{(8)}\,\hat{J}_{3},
\label{eq:mu0} \\
\hat{\mu}^{(1)} & =  
\frac{w_{4}}{\sqrt{3}}\,d_{pq3}D_{Q p}^{(8)}\,D_{8q}^{(8)}
+w_{5}\,\left(D_{Q3}^{(8)}\,D_{88}^{(8)} +
                  D_{Q8}^{(8)}\,D_{83}^{(8)}\right)  
+w_{6}\,\left(D_{Q3}^{(8)}\,D_{88}^{(8)} -
                  D_{Q8}^{(8)}\,D_{83}^{(8)}\right).  
\label{eq:mu1}
\end{align}
$\hat{g}_1^{(0)}$ and $\hat{\mu}^{(0)}$ denote the SU(3)
symmetric parts of the collective operators for the axial-vector and
magnetic dipole transition, respectively. $\hat{g}_1^{(1)}$ and
$\hat{\mu}^{(1)}$ represent those of the SU(3) symmetry breaking,
respectively. $d_{3bc}$ and $d_{pq3}$ designate the SU(3) symmetric
invariant tensors. The subscript index $\varphi$ of $D_{\varphi 8}^{(8)}$
in the axial-vector operator means a pseudoscalar meson in the final
state of a strong decay of $N_{\overline{\bm{10}}}$
($N_{\bm{27}}$). $Q$ of $D_{Q 8}^{(8)}$ in the magnetic dipole
transition operator stands for the electric charge $Q=T_3+Y/2$. The
dynamical parameters $a_i$ and $w_i$ encode specific dynamics of the
$\chi$QSM. $a_i$ were determined by using the experimental data on
hyperon semileptonic decays of the baryon octet and the empirical
value of the singlet axial-vector charge of the proton whereas $w_i$
were fixed by those on the magnetic moments of the baryon octet. We
refer to Ref.~\cite{Yang:2015era} for the details of how these
parameters were determined. Thus, we take the numerical values of
$a_i$ and $w_i$ from Ref.~\cite{Yang:2015era}:  
\begin{align}
a_1 &= -3.509\pm0.011, \;\;\;\; 
a_2 \;=\; 3.437\pm0.028,\;\;\;\;
a_3\;=\; 0.604\pm0.030,  
\cr
a_4&= -1.213\pm0.068, \;\;\;\; 
a_5 \;=\; 0.479\pm0.025,\;\;\;\;  
a_6\;=\; -0.735\pm0.040, 
\label{eq:ai} 
\end{align}
and
\begin{align}
w_1 &= -13.515\pm0.010, \;\;\;\; 
w_2 \;=\; 4.147\pm0.933,\;\;\;\; 
w_3\;=\; 8.544\pm0.861, 
\cr
w_4 & = -3.793\pm0.209, \;\;\;\; 
w_5 \;=\; -4.928\pm0.862,\;\;\;\;
w_6\;=\; -2.013\pm0.842. 
\label{eq:wi}
\end{align}
We want to mention that strong decay widths of the baryon decuplet 
have been reproduced in very good agreement with the experimental
data~\cite{Yang:2018idi} with the numerical values of $a_i$ in
Eq.~\eqref{eq:ai} employed. The magnetic moments of the baryon
decuplet were also obtained to be in good agreement with the
data. 

Sandwiching the axial-vector transition operator between the baryon
states given in Eq~\eqref{eq:su3wf}, we obtain the axial-vector
transition constants $g_{1}^{B_{i}\rightarrow B_{f}}$. 
The explicit expressions for the axial-vector transition constants
$g_{1}^{\left(B_{i}\rightarrow B_{f}\right)}$ can be found in
Appendix~\ref{app:c}. Since, however, the pseudovector coupling
constants, $f_{\varphi   B_{f}B_{i}}$, are often used in the
description of hadronic reaction, we will relate the axial-vector
transition constants to them by the usual formula 
\begin{align}
f_{\varphi B_{f}B_{i}}  =  
\frac{m_{\varphi}}{f_{\varphi}}g_{1}^{B_{i}\rightarrow B_{f}},
\label{eq:fg1}
\end{align}
where $m_{\varphi}$ and $f_{\varphi}$ denote the mass and the decay 
constant of a pseudoscalar meson involved in a strong decay,
respectively. To obtain the pseudovector coupling constants, we have
used the following numerical values of the pseudoscalar meson masses
and decay constants \cite{Behrend:1990sr, Tanabashi:2018oca}
\begin{align}
  \label{eq:inputf}
f_\pi &= 92.4\,\mathrm{MeV},\;\;\; m_\pi = 137.57\, \mathrm{MeV}, \cr
f_K &= 113.0\,\mathrm{MeV},\;\;\; m_K = 493.7\,\mathrm{MeV},\cr
f_\eta &=  94.0\,\mathrm{MeV},\;\;\; m_\eta = 547.9 \,\mathrm{MeV}.
\end{align}
Using the results of Eq.~\eqref{eq:fg1}, we can easily
derive the expression for the strong decays as follows
\begin{align}
\Gamma_{B_{i}\rightarrow\varphi+B_{f}} & =  
\frac{\left|\bm{P}_{\varphi}\right|^{3}}{6\pi m_{\varphi}^{2}}
\frac{M_{f}}{M_{i}}\left(f_{\varphi B_{f}B_{i}}\right)^{2},
\label{eq:width}
\end{align}
where $\bm{P}_{\varphi}$ is the momentum of the outgoing pseudoscalar
meson $\varphi$. Note that the strong coupling constants $f_{\varphi 
  B_{f}B_{i}}$ include the factors coming from the average and
summation over the initial and final spin and isospin states,
respectively. 

Since we are mainly interested in the decays of
$N^*(1685)1/2^+$ and $N^*(1726)3/2^+$ into $\eta N$, we need to derive the 
$\eta N N_{\overline{\bm{10}}} (N_{\bm{27}})$ coupling constants,
which means that we need to extract them from the singlet $\eta_0$ and
octet $\eta_8$ coupling constants. Introducing the
mixing angle, we can get $f_{\eta B_{f}B_{i}}$ and
$f_{\eta^{\prime}B_{f}B_{i}}$ as follows
\begin{align}
f_{\eta B_{f}B_{i}} & =  
\mathrm{cos}\theta_{p}\,f_{\eta_8 B_{f}B_{i}}
\;-\;\mathrm{sin}\theta_{p}\,f_{\eta_0 B_{f}B_{i}},\cr
f_{\eta^{\prime}B_{f}B_{i}} & =  
\mathrm{sin}\theta_{p}\,f_{\eta_8 B_{f}B_{i}}
\;+\;\mathrm{cos}\theta_{p}\,f_{\eta_0 B_{f}B_{i}},
\label{eq:etamixing}
\end{align}
where the mixing angle $\theta_{p}=-15.5{}^{\circ}$ is taken from
Ref. \cite{Bramon:1997va}. 

\begin{table}[htp]
\caption{Numerical results of the pseudovector coupling constants and
  strong decay widths of the antidecuplet nucleon and eikosiheptaplet
  nucleon with spin 3/2.}. 
\label{tab:2} %
\begin{tabular}{ccccccc}
\hline 
${\bm{\mathcal{R}}}_{J}$  
& $B_{i}\rightarrow\varphi+B_{f}$  
& $f_{\varphi B_{f}B_{i}}^{\left(0\right)}$  
& $f_{\varphi B_{f}B_{i}}^{\left(\mathrm{tot}\right)}$  
& $\Gamma_{\varphi
  B_{f}B_{i}}^{\left(0\right)}\left[\mathrm{MeV}\right]$  
& $\Gamma_{\varphi
  B_{f}B_{i}}^{\left(\mathrm{tot}\right)}\left[\mathrm{MeV}\right]$  
& $\Gamma_{\varphi
  B_{f}B_{i}}^{\left(\mathrm{Full}\right)}\left[\mathrm{MeV}\right]$
\cr
\hline 
\multirow{4}{*}{${\overline{\bm{10}}}_{1/2}$}  
& $N_{\overline{\bm{10}}}\rightarrow\pi+N$  
& $-0.04\pm0.01$  
& $-0.17\pm0.01$  
& $0.42\pm0.12$  
& $8.34\pm1.03$  
& \multirow{4}{*}{$30.5\pm5.0$}
\cr
 & $N_{\overline{\bm{10}}}\rightarrow\eta+N$  
& $0.96\pm0.11$  
& $1.95\pm0.20$  
& $5.37\pm1.31$  
& $22.09\pm4.89$  
& \cr
 & $N_{\overline{\bm{10}}}\rightarrow K+\Lambda$  
& $-0.11\pm0.02$  
& $-0.16\pm0.02$  
& $0.02\pm0.01$  
& $0.05\pm0.02$  
& \cr
 & $N_{\overline{\bm{10}}}\rightarrow K+\Sigma$  
& $-0.11\pm0.02$  
& $0.055\pm0.024$  
& $0.0001\pm0.0008$  
& $\sim0.00003$  
& \cr
\hline 
\multirow{6}{*}{${\bm{27}}_{3/2}$}  
& $N_{{\bm{27}}}\rightarrow\pi+N$  
& $-0.11\pm0.01$  
& $0.04\pm0.01$  
& $4.2\pm0.1$  
& $0.6\pm0.1$  
& \multirow{4}{*}{$22.2\pm6.2$}
\cr
 & $N_{{\bm{27}}}\rightarrow\eta+N$  
& $0.43\pm0.18$  
& $1.61\pm0.27$  
& $1.3\pm1.1$  
& $18.7\pm6.2$  
& \cr
 & $N_{{\bm{27}}}\rightarrow K+\Lambda$  
& $-1.01\pm0.02$  
& $-0.93\pm0.02$  
& $3.2\pm0.3$  
& $2.7\pm0.3$  
& \cr
 & $N_{{\bm{27}}}\rightarrow K+\Sigma$  
& $0.34\pm0.01$  
&
 $0.61\pm0.02$  
& $0.06\pm0.02$  
& $0.19\pm0.07$  
& \cr
\hline 
\end{tabular}
\end{table}
The numerical results of the pseudovector coupling constants and the
strong decay widths for the $N^*(1685)1/2^+$ and $N^*(1726)3/2^+$ are
listed in Table~\ref{tab:2}. As shown from Table~\ref{tab:2}, the
leading-order contributions are suppressed, so that
the effects of SU(3) symmetry breaking become very important. This can
be understood by examining the leading-order expressions of the
axial-vector coupling constants for the vertices
$N_{\overline{\bm{10}}}\to \varphi +N$ and $N_{\bm{27}}\to \varphi
+N$ in Eqs.~\eqref{eq:C1}-\eqref{eq:C8}. The expressions for
$g_1^{(0)} [N_{\overline{\bm{10}}} \to \varphi + N]$ are all
proportional to $a_1+a_2+a_3/2$. As given in Eq.~\eqref{eq:ai}, the
values of $a_1$ and $a_2$ are almost the same but the signs are
different each other. On the other hand, that of $a_3$ is rather
small. Consequently, the numerical results of $g_1^{(0)}
[N_{\overline{\bm{10}}} \to \varphi + N]$ turn out to be very small.  
The formulas for $g_1^{(0)} [N_{\bm{27}} \to \varphi + N]$ are
similarly proportional to $a_1+a_2/2$, which also brings about the
suppression of the leading-order contribution to the axial-vector
transition coupling constants of $N_{\bm{27}}$. Thus, the
contributions of the SU(3) symmetry breaking come into play of 
leading roles. We want to emphasize that, however, the situation is
opposite when it comes to the case of the $B_{\bm{10}}\to B_{\bm{8}}$
transitions. The leading-order expressions for $g_1^{(0)}
[B_{\bm{10}}\to \varphi + B_{\bm{8}}]$ are all proportional to $a_1
-a_2/2$, which causes indeed the leading-order contributions to be the
most dominant ones.   

The results presented in Table~\ref{tab:2} have important physical
implications. The strong decay widths of both 
$N_{\overline{\bm{10}}}\to \eta + N$ and $N_{\bm{27}}\to \eta+N$ are
much larger than those of $N_{\overline{\bm{10}}}\to \pi + N$ and
$N_{\bm{27}}\to \pi+N$, respectively: the value of 
$\Gamma[N_{\overline{\bm{10}}}\to \eta + N]$ is approximately 2.6
times larger than that of $\Gamma[N_{\overline{\bm{10}}}\to \pi + N]$
and that of $\Gamma[N_{\bm{27}}\to \eta + N]$ is even 31 times larger
than that of $\Gamma[N_{\bm{27}}\to \pi + N]$. This explains why both
the narrow resonances $N^*(1685)$ and $N^*(1726)$ are more likely to be
observed in $\eta$ photoproduction than in $\gamma+N\to \pi+N$. 
The full decay widths of $N_{\overline{\bm{10}}}$ and $N_{\bm{27}}$
are obtained respectively as $(30.5\pm5.0)$ MeV and $(22.2\pm 6.2)$
MeV, which are indeed much narrower than usual excited nucleon
resonances. In particular, the strong decay width of $N^*(1685)$ is in
remarkable agreement with the experimental data on the
corresponding intrinsic width that was estimated to be $(30\pm 15)$
MeV~\cite{Kuznetsov:2006kt, Jaegle:2011sw, Werthmuller:2013rba,
  Witthauer:2013tkm}. Note that the strong decay width of
$N^*(1726)$ is predicted to be even narrower than that of $N^*(1685)$. 
However, there is one caveat. The present work shows relatively larger
value of $\Gamma[N_{\overline{\bm{10}}}\to \pi + N]$, though it is
still quite smaller than that of $\Gamma[N_{\overline{\bm{10}}}\to
\eta + N]$. As pointed out by Goeke et al.~\cite{Goeke:2009ae}, the
mixing of the $N_{\overline{\bm{10}}}$ with the Roper resonance
provides a crucial explanation of why the $N_{\overline{\bm{10}}}$ was
not seen in the $\pi N$ scattering data. Thus, the inclusion of the
mixing with $N(1440)$ will further decrease the value of
$\Gamma[N_{\overline{\bm{10}}}\to \pi + N]$. However, in order to
investigate the effects of this mixing, one needs to construct first a 
formalism of describing the spectra of excited baryons within the same
framework as the present one. We leave it as a future work. 

It is also interesting to look into the decay modes of both
$N_{\overline{\bm{10}}}$ and $N_{\bm{27}}$ with strangeness. As
displayed in Table~\ref{tab:2}, the decay of
$N_{\overline{\bm{10}}}\to K+\Sigma$ is almost forbidden, because the
mass of $N_{\overline{\bm{10}}}$ is smaller than the threshold of the
$K\Sigma$ production. The decay of $N_{\overline{\bm{10}}}\to
K+\Lambda^0$ is allowed but the magnitude of the corresponding decay 
width is rather tiny. However, recent experiments on $K \Lambda^0$
photoproduction off the quasi-free neutron~\cite{Tsuchikawa:2016ixc,
  Tsuchikawa:2017tqm, Compton:2017xkt,Ho:2018riy} provide some hint on
the existence of the narrow nucleon resonance~\cite{Kim:2018qfu}.  In
contrast, the eikosiheptaplet nucleon can decay into $K$ and $\Sigma$,
though the partial decay width is rather small. However, the 
width for the $N_{\bm{27}}\to K+\Lambda$ is about $(2.7\pm
0.3)\,\mathrm{MeV}$, which might be detectable experimentally.   

\begin{table}[htp]
\caption{Magnetic transition moments in units of the nuclear magneton
  ($\mu_N$) and radiative decay widths of the antidecuplet
and eikosiheptaplet nucleons in units of keV.} 
\label{tab:3} %
\begin{tabular}{cccc}
\hline 
$N_{\overline{\bm{10}}}\rightarrow\gamma+N$  
& $\mu_{N_{\overline{\bm{10}}} N}^{\left(0\right)}$  
& $\mu_{N_{\overline{\bm{10}}} N}^{\left(\mathrm{tot}\right)}$  
& $\Gamma_{N_{\overline{\bm{10}}} N
  \gamma}^{\left(\mathrm{tot}\right)} 
\left[\mathrm{keV}\right]$ 
\cr
\hline 
$p_{\overline{\bm{10}}}\rightarrow\gamma+p$  
& $0$  
& $0.15\pm0.04$  
& $18.78\pm0.52$ 
\cr
$n_{\overline{\bm{10}}}\rightarrow\gamma+n$  
& $-0.38\pm0.08$  
& $-0.44\pm0.09$  
& $161.83\pm64.72$ 
\cr
\hline 
$N_{{\bm{27}}}\rightarrow \gamma+N$  
& $\mu_{{\bm{27}}}^{\left(0\right)}$  
& $\mu_{{\bm{27}}}^{\left(\mathrm{tot}\right)}$  
& $\Gamma_{\gamma N_{\bm{27}} N}^{\left(\mathrm{tot}\right)}
  \left[\mathrm{MeV}\right]$  
\cr
\hline 
$p_{{\bm{27}}}\rightarrow\gamma+p$  
& $-0.93\pm0.04$  
& $-0.75\pm0.05$  
& $1.43\pm0.19$ 
\cr
$n_{{\bm{27}}}\rightarrow\gamma+n$  
& $-0.46\pm0.02$  
& $-0.38\pm0.02$  
& $0.38\pm0.04$ 
\cr
\hline 
\end{tabular}
\end{table}
As mentioned in Introduction, the magnetic dipole transitions of the
$N_{\overline{\bm{10}}}$ have been investigated qualitatively in
previous works~\cite{Polyakov:2003dx, Kim:2005gz} in which a
theoretical ambiguity was unavoidable because the baryon wavefunctions
could not be fixed when the effects of SU(3) symmetry breaking were
included. This uncertainty led to the fact that the magnetic 
transition moments of $N_{\overline{\bm{10}}}$ are proportional to
$\Sigma_{\pi N}$. Unfortunately, the leading-order contributions to
$\mu_{NN_{\overline{\bm{10}}}}$ were very sensitive to the change of
the $\Sigma_{\pi N}$ value, so that a precise prediction was not
possible in Ref.~\cite{Kim:2005gz}. However, the present work does not
have such an ambiguity anymore. In Table~\ref{tab:3}, we list the
numerical results of the magnetic transition moments of both 
$N_{\overline{\bm{10}}}$ and $N_{\bm{27}}$. As pointed out by
Refs.~\cite{Polyakov:2003dx, Kim:2005gz}, the leading-order
contributions to the magnetic transition moments of
$N_{\overline{\bm{10}}}$ are proportional to $Q-1$, where $Q$ denotes
the corresponding charge of $N_{\overline{\bm{10}}}$. Thus,
$\mu[p_{\overline{\bm{10}}}\to \gamma + p]$ exactly vanishes in the
leading order whereas $\mu[n_{\overline{\bm{10}}}\to \gamma + n]$
has a finite value. The situation is the other way around in the case
of $N_{\bm{27}}$: those of $N_{\bm{27}}$ are proportional to $Q+1$, so
that the leading-order value of $\mu[p_{\bm{27}}\to \gamma + p]$ turns
out to be larger than that of $\mu[n_{\bm{27}}\to \gamma + n]$, which
is opposite to the case of $N_{\overline{\bm{10}}}$. 

The expressions of the magnetic transition
moments for the $N_{\overline{\bm{10}}}$  and $N_{\bm{27}}$ are very
similar to those of the axial-vector transitions. Namely, the
leading-order contributions to $\mu[N_{\overline{\bm{10}}}\to \gamma +
N]$ are proportional to $w_1+w_2+w_3/2$ whereas  $\mu[N_{\bm{27}}\to
\gamma + N]$ is in proportion to $w_1+w_2/2$. It explains again the
reason why the magnetic transition moments of 
$n_{\overline{\bm{10}}}$ and $N_{\bm{27}}$ are rather small in
leading order. The contributions of the SU(3) symmetry breaking become
the leading one to $\mu[p_{\overline{\bm{10}}}\to \gamma + p]$. In the
case of $\mu[n_{\overline{\bm{10}}}\to \gamma + n]$, the effects of
the SU(3) symmetry breaking contribute to
$\mu[N_{\overline{\bm{10}}}\to \gamma +N]$ approximately by $25\,\%$ 
and to $\mu[N_{\bm{27}}\to \gamma +N]$ by about $20\,\%$. 

By the reason explained above, the radiative decay width of
the antidecuplet neutron turns out to be much larger than that of the
antidecuplet proton. Their ratio can be explicitly obtained as 
\begin{align}
\frac{\Gamma_{\gamma} \left[n_{\overline{\bm{10}}} \rightarrow n\right] }{
  \Gamma_{\gamma}\left[p_{\overline{\bm{10}}} \rightarrow p \right]}  
 =  8.62\pm3.45.
\label{eq:ratioNstar}
\end{align}
Thus, the neutron anomaly can be explained by this ratio, as already
pointed out by Refs.~\cite{Polyakov:2003dx, Kim:2005gz}. 
However, when it comes to the radiative decays of the eikosiheptaplet
nucleons, $p_{\bm{27}}$ has a larger radiative decay width than
$n_{\bm{27}}$ does. Thus, their ratio is obtained as 
\begin{align}
\frac{\Gamma_{\gamma} \left[p_{{\bm{27}}}\rightarrow p \right]}{
  \Gamma_{\gamma} \left[n_{{\bm{27}}} \rightarrow n \right]}  
& =  3.76\pm0.64.
\label{eq:ratioN27}
\end{align}
It indicates that the eikosiheptaplet nucleon is more likely to be
found in $\eta$ photoproduction off the proton. It is of great
interest if we have a proton anomaly in finding $N_{\bm{27}}$ though
it is not as prominent as the neutron anomaly in
$N_{\overline{\bm{10}}}$. 

\section{Summary and conclusions
\label{sec:Conclusion}} 
In the present work, we have investigate the strong and radiative
decay widths of the antidecuplet and eikosiheptaplet nucleons in
addition to their masses, based on the SU(3) chiral quark-soliton
model. All the relevant parameters for the strong and radiative decay
widths have been already fixed in the baryon octet sector, so that we
do not have any additional parameter to obtain the numerical results of
the strong and radiative decay widths. From the present study, we come
to the following conclusions:
\begin{itemize}
\item The second narrow peak found in Refs.~\cite{Gridnev:2016dba,
    Kuznetsov:2017xgu, Werthmuller:2015owc} can be identified 
  as a member of the eikosiheptaplet with spin
  3/2~\cite{Praszalowicz:2007zza}.  Thus, the quantum numbers of this
  narrow nucleon resonance will be given as spin 3/2 and negative
  parity, i.e. $N^*(1726)3/2^+$. Other members of the eikosiheptaplet
  with spin 1/2 have rather large masses,
  i.e. $M_{N_{\bm{27}(J=1/2)}}>2\, \mathrm{GeV}$. Thus, we did not
    discuss them in this work. We have predicted the mass of
    $N^*(1726)3/2^+$ to be $M_{N_{\bm{27}}}=(1719.6\pm7.4)$ MeV.  
\item The partial decay width $\Gamma_{N_{\overline{\bm{10}}}\to \eta
    N}$ of the antidecuplet nucleon is at least three times larger
  than $\Gamma_{N_{\overline{\bm{10}}}\to \pi
    N}$. $\Gamma_{N_{\bm{27}} \to \eta N}$ is even 31 times larger
  than $\Gamma_{N_{\bm{27}} \to \pi N}$. Note, however, that
  Gridnev et al.~\cite{Gridnev:2016dba}  have seen both the narrow
  resonant  structures from the analysis of $\pi p$ elastic
  scattering. The present results imply that the narrow nucleon
  resonance $N^*(1726)3/2^+$ may be found in hadronic or photonic
  processes with the $\eta$ meson involved. 
\item We found the ratio of the radiative decay widths for
  $N^*(1685)1/2^+$    $\Gamma_{n_{\overline{\bm{10}}}(1685)\,n} / 
  \Gamma_{p_{\overline{\bm{10}}}(1685)\,p} = 8.62$. This explains the
  reason for the neutron anomaly as already pointed out in
  Refs.~\cite{Polyakov:2003dx, Kim:2005gz}. Note that $
  \Gamma_{p_{\overline{\bm{10}}}(1685)\,p}$ vanishes in the limit of
  SU(3) symmetry. Thus, the contribution only arises from the effects
  of the SU(3) symmetry breaking. It leads to the large ratio of
  $\Gamma_{n_{\overline{\bm{10}}}(1685)\,n} /
  \Gamma_{p_{\overline{\bm{10}}}(1685)\,p}$. 
\item On the other hand, the ratio of the radiative decay widths for
  $N^*(1726)3/2^+$ was found to be
  $\Gamma_{p_{\bm{27}}(1726)\,p}/\Gamma_{n_{\bm{27}}(1726)\,n} = 
  3.76$, which is opposite to the case of $N_{\overline{\bm{10}}}$.  
Though the size of this ratio is not as noticeable as that for 
$N_{\overline{\bm{10}}}$, the eikosiheptaplet proton is more likely to
be observed in comparison with the corresponding neutron. If it is
experimentally true, we can call it \textit{proton anomaly}. 
\item  Last but not least, it is of great interest to examine the
  decay modes of the antidecuplet and eikosiheptaplet nucleons with
  strangeness. As expected, $N_{\overline{\bm{10}}}$ does not decay
  into $K$ and $\Sigma$ on account of the fact that the corresponding
  threshold energy is higher than the mass of
  $N_{\overline{\bm{10}}}$. The decay width for
  $N_{\overline{\bm{10}}}\to K + \Lambda$ turns out to be very small. 
In contrast to the antidecuplet nucleons, $N_{\bm{27}}$ is allowed to
decay into $K$ and $\Sigma$, though its magnitude is rather small. The
decay width for $N_{\bm{27}}\to K + \Lambda$ is $(2.7\pm
0.3)\,\mathrm{MeV}$, which might be observed in $K\Lambda^0$
photoproduction off the proton.
\end{itemize}
\section*{Acknowledgment}
H.-Ch. K is grateful to A. Hosaka, T. Maruyama, M. Oka for useful
discussions.  He wants to express his gratitude to the members of the 
Advanced Science Research Center, Japan Atomic Energy Agency, 
where part of the present work was done. Gh.-S. Y is also grateful to
H.D. Son for valuable discussions. The present work was supported by
Basic Science Research Program through the National Research
Foundation of Korea funded by the Ministry of Education, Science and
Technology (No. NRF-2019R1A2C1010443 (Gh.-S. Y.) and 2018R1A5A1025563 
(H.-Ch.K.)).   
\begin{appendix}

\section{Mixing coefficients in Eq.~\eqref{eq:su3wf}}
\label{sec:Appendix1}
In this Appendix, we present the explicit expressions for the mixing
coefficients in Eq.~\eqref{eq:su3wf}:
\begin{align}
c_{\overline{\bm{10}}}^{B}  =  -\frac{1}{3}I_{2}\Delta_{\mathrm{s}}
\left(\alpha+\frac{1}{2}\gamma\right)
\left(\begin{array}{ccc}
{\overline{\bm{10}}} & {\bm{8}} & {\bm{8}}\cr
B & 0,0,0 & B
\end{array}\right),\;\;\;\;
c_{\bm{27}}^{B} =  \frac{3}{5\sqrt{5}}I_{2} \Delta_{\mathrm{s}}
\left(\alpha-\frac{1}{6}\gamma\right)
\left(\begin{array}{ccc}
\bf 27 & \bf 8 & \bf 8
\cr
B & 0,0,0 & B
\end{array}\right),
\label{eq:mixc}
\end{align}

\begin{align}
a_{{\bm{27}}}^{B}  = 
-\frac{3}{4}I_{2}  \Delta_{\mathrm{s}}
\left(\alpha+\frac{5}{6}\gamma\right)
\left(\begin{array}{ccc}
\bf 27 &\bf  8 &\bf  10\cr
B & 0,0,0 & B
\end{array}\right),\;\;\;\;
a_{\bm{35}}^{B}  = 
\frac{5}{12\sqrt{5}}I_{2}  \Delta_{\mathrm{s}}
\left(\alpha-\frac{1}{2}\gamma\right)
\left(\begin{array}{ccc}
\bf 35 &\bf  8 &\bf  10\cr
B & 0,0,0 & B
\end{array}\right),
\label{eq:mixa}
\end{align}

\begin{align}
d_{\bm{8}}^{B} & = \frac{2}{3\sqrt{5}}I_{2} \Delta_{\mathrm{s}}
\left(\alpha+\frac{1}{2}\gamma\right)
\left(\begin{array}{ccc}
\bf 8 &\bf  8 & {\overline{\bm{10}}}\cr
B & 0,0,0 & B
\end{array}\right),\;\;\;\;
d_{{\bm{27}}}^{B}  = \frac{3}{4\sqrt{5}}I_{2} \Delta_{\mathrm{s}}
\left(\alpha-\frac{7}{6}\gamma\right)
\left(\begin{array}{ccc}
\bf 27 &\bf  8 & {\overline{\bm{10}}}\cr
B & 0,0,0 & B
\end{array}\right),\cr
d_{\overline{\bm{35}}}^{B} 
& = 
\frac{1}{4}I_{2} \Delta_{\mathrm{s}}
\left(\alpha+\frac{1}{6}\gamma\right)
\left(\begin{array}{ccc}
\bf \overline{35} & \bf 8 & {\overline{\bm{10}}}\cr
B & 0,0,0 & B
\end{array}\right),
\label{eq:mixd}
\end{align}

\begin{align}
n_{\bm{8}}^{B} & = \frac{4}{5\sqrt{30}}I_{2} \Delta_{\mathrm{s}}
\left(\alpha-\frac{1}{6}\gamma\right)
\left(\begin{array}{ccc}
\bf 8 &\bf  8 & {\bm{27}}\cr
B & 0,0,0 & B
\end{array}\right),\;\;\;\;
n_{\overline{\bm{10}}}^{B}  = \frac{1}{2\sqrt{6}}I_{2}\Delta_{\mathrm{s}}
\left(\alpha-\frac{7}{6}\gamma\right)
\left(\begin{array}{ccc}
{\overline{\bm{10}}} &\bf  8 & {\bm{27}}\cr
B & 0,0,0 & B
\end{array}\right),\cr
n_{\overline{\bm{35}}}^{B} & = -\frac{5}{8\sqrt{15}}I_{2}\Delta_{\mathrm{s}}
\left(\alpha+\frac{5}{6}\gamma\right)
\left(\begin{array}{ccc}
\bf \overline{35} &\bf  8 & {\bm{27}}\cr
B & 0,0,0 & B
\end{array}\right),\;\;\;
n_{\bm{64}}^{B}  = \frac{20}{7\sqrt{210}}I_{2} \Delta_{\mathrm{s}}
\left(\alpha-\frac{1}{6}\gamma\right)
\left(\begin{array}{ccc}
\bf 64 &\bf 8 & {\bm{27}}\cr
B & 0,0,0 & B
\end{array}\right),
\label{eq:mixn}
\end{align}

\begin{align}
m_{\bm{10}}^{B} & = \frac{5}{2\sqrt{30}}I_{2} \Delta_{\mathrm{s}}
\left(\alpha+\frac{5}{6}\gamma\right)
\left(\begin{array}{ccc}
\bf 10 & \bf 8 & {\bm{27}}\cr
B & 0,0,0 & B
\end{array}\right),\;\;\;\;
m_{\bm{35}}^{B}  = 
\frac{5}{8\sqrt{15}}I_{2}\Delta_{\mathrm{s}}
\left(\alpha-\frac{7}{6}\gamma\right)
\left(\begin{array}{ccc}
\bf 35 & \bf 8 & {\bm{27}}\cr
B & 0,0,0 & B
\end{array}\right),\cr
m_{\overline{\bm{35}}}^{B}  &= -\frac{1}{2\sqrt{3}}I_{2}\Delta_{\mathrm{s}}
\left(\alpha+\frac{5}{6}\gamma\right)
\left(\begin{array}{ccc}
\bf \overline{35} & \bf 8 & {\bm{27}}\cr
B & 0,0,0 & B
\end{array}\right),\;\;\;\;
m_{\bm{64}}^{B}  = 
\frac{10}{7\sqrt{105}}I_{2}\Delta_{\mathrm{s}}
\left(\alpha-\frac{1}{6}\gamma\right)
\left(\begin{array}{ccc}
\bf 64 &\bf  8 & {\bm{27}}\cr
B & 0,0,0 & B
\end{array}\right).
\label{eq:mixm}
\end{align}
\section{Expressions for the masses of the 
antidecuplet and eikosiheptaplet nucleons}
\label{app:b}
In Table~\ref{tab:4}, we tabulate each contribution to the masses of
the eikosiheptaplet nucleons.  
\begin{table}[htp]
\global\long\def\arraystretch{1.5}
\caption{Expressions for the masses for the eikosiheptaplet nucleons}
\label{tab:4}
\begin{tabular}{c|ccrccc}
\hline 
\multicolumn{3}{c}{States} 
& {$T_{3}\;\;$}  
& EM  
& Isospin  
& $\mathrm{SU}_{f}(3)$ 
\cr
\hline 
\multirow{2}{*}{${\bm{27}}_{1/2}$}  
& \multirow{2}{*}{{$N_{{\bm{27}}}$}}  
& {$p_{{\bm{27}}}$}  
& {$1/2$}  
& ${\frac{33}{280}\left(\delta^{(8)}+\frac{6}{11}\delta^{(27)}\right)}$  
&
${-\frac{71}{1120} \Delta_{\mathrm{du}}
\left(\alpha-\frac{560}{71}\beta+\frac{233}{142}\gamma\right)}$  
&
\multirow{2}{*}{${\frac{137}{560} \Delta_{\mathrm{s}}
\left(\alpha+\frac{560}{137}\beta+\frac{71}{274}\gamma\right)}$} 
\cr
 &  & 
{$n_{{\bm{27}}}$}  
& {$-1/2$}  
& ${\frac{13}{35}\left(\delta^{(8)} +\frac{41}{156} \delta^{(27)} \right)}$  
&
${\frac{71}{1120} \Delta_{\mathrm{du}}
\left(\alpha-\frac{560}{71}\beta+\frac{233}{142}\gamma\right)}$  
& \cr
\hline 
\multicolumn{4}{c}{\vspace{-1.8em}} 
&  
&  
& \cr
\hline 
\multirow{2}{*}{${\bm{27}}_{3/2}$}  
& \multirow{2}{*}{{$N_{{\bm{27}}}$}}  
& {$p_{{\bm{27}}}$}  
& {$1/2$}  
& ${\frac{3}{14} \left(\delta^{(8)} -\frac{1}{4} \delta^{(27)} \right)}$  
&
${\frac{5}{56} \Delta_{\mathrm{du}}
\left(\alpha+\frac{28}{5}\beta-\frac{5}{2}\gamma\right)}$  
&
\multirow{2}{*}
{${\frac{1}{28} \Delta_{\mathrm{s}}
\left(\alpha+28\beta-\frac{5}{2}\gamma\right)}$} 
\cr
 &  & 
{$n_{{\bm{27}}}$}  
& {$-1/2$}  
&
 ${-\frac{1}{7} \left(\delta^{(8)} +\frac{19}{24} \delta^{(27)} \right)}$  
&
 ${-\frac{5}{56} \Delta_{\mathrm{du}}
\left(\alpha+\frac{28}{5}\beta-\frac{5}{2}\gamma\right)}$  
& \cr
\hline 
\end{tabular}
\end{table}
\section{Expressions for the axial-vector transition 
constants}
\label{app:c}
Expressions of axial-vector coupling constants of the baryon antidecuplet
$N_{\overline{\bm{10}}}$ to the baryon octet are given
below. $g_{1}^{\left(0\right)}$ denote the leading-order contributions
to the axial-vector transition constants. Note, however, that it
contains the rotational $1/N_c$ corrections. We will not decompose
them in the present work and call it generically the leading-order
terms. $g_{1}^{\left(\mathrm{op}\right)}$ represent the contributions
of the SU(3) symmetry breaking arising from the linear
$m_{\mathrm{s}}$ expansion 
given in Eq.~\eqref{eq:gmu}. $g_{1}^{\left(\mathrm{wf}\right)}$ stand
for those from the baryon wavefunctions that also contains the linear
$m_{\mathrm{s}}$. 
\begin{align}
g_{1}^{\left(0\right)}\left[N_{\overline{\bm{10}}}\rightarrow\pi+N\right] 
& =  -\frac{1}{6\sqrt{5}}\left(a_{1}+a_{2}+\frac{1}{2}a_{3}\right),
\cr
g_{1}^{\left(\mathrm{op}\right)}\left[N_{\overline{\bm{10}}}\rightarrow\pi+N\right] 
& =  -\frac{1}{54\sqrt{5}}\left(a_{4}+6a_{5}+9a_{6}\right),
\cr
g_{1}^{\left(\mathrm{wf}\right)}\left[N_{\overline{\bm{10}}}\rightarrow\pi+N\right] 
& = 
-\frac{5}{24\sqrt{5}}\left(a_{1}+\frac{5}{2}a_{2}-\frac{1}{2}a_{3}\right)
      c_{\overline{\bm{10}}}
\;-\;\frac{49}{72\sqrt{5}}\left(a_{1}-\frac{11}{14}a_{2}-\frac{3}{14}a_{3}\right)
      c_{\bm{27}} 
\cr
 &  
 -\frac{7}{6\sqrt{5}}\left(a_{1}-\frac{1}{2}a_{2}-\frac{1}{14}a_{3}\right)
      d_ {\bm{8}}
\;-\;\frac{1}{90\sqrt{5}}\left(a_{1}+2a_{2}-\frac{3}{2}a_{3}\right)
      d_{{\bm{27}}}, 
\label{eq:C1}
\end{align}
\begin{align}
g_{1}^{\left(0\right)}\left[N_{\overline{\bm{10}}}\rightarrow\eta+N\right] 
& =  \frac{1}{2\sqrt{15}}\left(a_{1}+a_{2}+\frac{1}{2}a_{3}\right),\cr
g_{1}^{\left(\mathrm{op}\right)}\left[N_{\overline{\bm{10}}}\rightarrow\eta+N\right] 
& =  -\frac{1}{6\sqrt{15}}a_{4},\cr
g_{1}^{\left(\mathrm{wf}\right)}\left[N_{\overline{\bm{10}}}\rightarrow\eta+N\right] 
& =  0,
\label{eq:C2}
\end{align}
\begin{align}
g_{1}^{\left(0\right)}\left[N_{\overline{\bm{10}}}\rightarrow K+\Lambda\right] 
& =  
-\frac{1}{2\sqrt{15}}\left(a_{1}+a_{2}+\frac{1}{2}a_{3}\right),\cr
g_{1}^{\left(\mathrm{op}\right)}\left[N_{\overline{\bm{10}}}\rightarrow
  K+\Lambda\right] 
& =  
\frac{1}{12\sqrt{15}}\left(a_{4}+3a_{6}\right),\cr
g_{1}^{\left(\mathrm{wf}\right)}\left[N_{\overline{\bm{10}}}\rightarrow
  K+\Lambda\right] 
& =  
\frac{7}{4\sqrt{15}}\left(a_{1}-\frac{11}{14}a_{2}-\frac{3}{14}a_{3}\right)c_{\bm{27}}
\;+\;\frac{2}{\sqrt{15}}\left(a_{1}-\frac{1}{2}a_{2}-\frac{1}{4}a_{3}\right)d_{\bm{8}}
\cr
 &  
 -\frac{1}{10\sqrt{15}}\left(a_{1}+2a_{2}-\frac{3}{2}a_{3}\right)d_{{\bm{27}}},
\label{eq:C3}
\end{align}
\begin{eqnarray}
g_{1}^{\left(0\right)}\left[N_{\overline{\bm{10}}}\rightarrow K+\Sigma\right] 
& = & 
-\frac{1}{6\sqrt{5}}\left(a_{1}+a_{2}+\frac{1}{2}a_{3}\right),
\cr
g_{1}^{\left(\mathrm{op}\right)}\left[N_{\overline{\bm{10}}}\rightarrow
  K+\Sigma\right] 
& = & 
-\frac{1}{108\sqrt{5}}\left(a_{4}-12a_{5}+9a_{6}\right),
\cr
g_{1}^{\left(\mathrm{wf}\right)}\left[N_{\overline{\bm{10}}}\rightarrow
  K+\Sigma\right] 
& = & 
-\frac{5}{12\sqrt{5}}\left(a_{1}+\frac{5}{2}a_{2}-\frac{1}{2}a_{3}\right)c_{\overline{\bm{10}}}
\;+\;\frac{7}{18\sqrt{5}}\left(a_{1}-\frac{11}{14}a_{2}-\frac{3}{14}a_{3}\right)c_{\bm{27}}
\cr
 &  &
 +\frac{1}{3\sqrt{5}}\left(a_{1}-\frac{1}{2}a_{2}+a_{3}\right)d_{\bm{8}}
\;+\;\frac{1}{90\sqrt{5}}\left(a_{1}+2a_{2}-\frac{3}{2}a_{3}\right)d_{{\bm{27}}}.
\label{eq:C4}
\end{eqnarray}
The following are the expressions of the axial-vector coupling
constants of the baryon eikosiheptaplet
$N_{{\bm{27}}}\left(J=3/2\right)$  to the baryon octet:  
\begin{align}
g_{1}^{\left(0\right)}\left[N_{{\bm{27}}}\rightarrow\pi+N\right] 
& =  \frac{2}{9\sqrt{30}}\left(a_{1}+\frac{1}{2}a_{2}\right),
\cr
g_{1}^{\left(\mathrm{op}\right)}\left[N_{{\bm{27}}}\rightarrow\pi+N\right] 
& =  \frac{2}{63\sqrt{30}}\left(a_{4}+4a_{5}-\frac{7}{2}a_{6}\right),
\cr
g_{1}^{\left(\mathrm{wf}\right)}\left[N_{{\bm{27}}}\rightarrow\pi+N\right] 
& =  -\frac{35}{18\sqrt{30}}\left(a_{1}-a_{2}\right)c_{\overline{\bm{10}}}
\;+\;\frac{191}{14\sqrt{30}}\left(a_{1}-\frac{8}{19}a_{2}\right)c_{\bm{27}},
\label{eq:C5}
\end{align}
\begin{align}
g_{1}^{\left(0\right)}\left[N_{{\bm{27}}}\rightarrow\eta+N\right] 
& =  \frac{2}{3\sqrt{10}}\left(a_{1}+\frac{1}{2}a_{2}\right),
\cr
g_{1}^{\left(\mathrm{op}\right)}\left[N_{{\bm{27}}}\rightarrow\eta+N\right] 
& =  -\frac{5}{21\sqrt{10}}\left(a_{4}-\frac{1}{5}a_{5}\right),
\cr
g_{1}^{\left(\mathrm{wf}\right)}\left[N_{{\bm{27}}}\rightarrow\eta+N\right] 
& =  0,
\label{eq:C6}
\end{align}
\begin{align}
g_{1}^{\left(0\right)}\left[N_{{\bm{27}}}\rightarrow K+\Lambda\right] 
& =  \frac{2}{3\sqrt{10}}\left(a_{1}+\frac{1}{2}a_{2}\right),
\cr
g_{1}^{\left(\mathrm{op}\right)}\left[N_{{\bm{27}}}\rightarrow
  K+\Lambda\right] 
& = 
-\frac{1}{14\sqrt{10}}\left(a_{4}+\frac{5}{3}a_{5}-\frac{7}{3}a_{6}\right),
\cr
g_{1}^{\left(\mathrm{wf}\right)}\left[N_{{\bm{27}}}\rightarrow
  K+\Lambda\right] 
& =  -\frac{3}{7\sqrt{10}}\left(a_{1}-\frac{16}{3}a_{2}\right)c_{\bm{27}},
\label{eq:C7}
\end{align}
\begin{align}
g_{1}^{\left(0\right)}\left[N_{{\bm{27}}}\rightarrow K+\Sigma\right] 
& =  -\frac{2}{9\sqrt{30}}\left(a_{1}+\frac{1}{2}a_{2}\right),
\cr
g_{1}^{\left(\mathrm{op}\right)}\left[N_{{\bm{27}}}\rightarrow
  K+\Sigma\right] 
& = 
-\frac{11}{126\sqrt{30}}\left(a_{4}-\frac{47}{11}a_{5}+\frac{49}{11}a_{6}\right),
\cr
g_{1}^{\left(\mathrm{wf}\right)}\left[N_{{\bm{27}}}\rightarrow
  K+\Sigma\right] 
& = 
-\frac{5}{9\sqrt{30}}\left(a_{1}-a_{2}\right)c_{\overline{\bm{10}}}
\;+\;\frac{6}{7\sqrt{30}}\left(a_{1}+\frac{22}{9}a_{2}\right)c_{\bm{27}}.
\label{eq:C8}
\end{align}
\section{Expressions for the dipole magnetic 
transition moments} 
\label{app:d}
Expressions of axial-vector coupling constants of the baryon antidecuplet
$N_{\overline{\bm{10}}}$ to the baryon octet are given
below (see also
Ref.~\cite{Kim:2005gz}). $\mu_{NN_{\overline{\bm{10}}}}^{\left(0\right)}$ 
denote the leading-order contributions 
to the axial-vector transition constants. Note, however, that it
contains the rotational $1/N_c$ corrections. We will not decompose
them in the present work and call it generically the leading-order
terms. $\mu_{NN_{\overline{\bm{10}}}}^{\left(\mathrm{op}\right)}$
represent the contributions of the SU(3) symmetry breaking arising
from the linear $m_{\mathrm{s}}$ expansion  
given in Eq.~\eqref{eq:mu}. 
$\mu_{NN_{\overline{\bm{10}}}}^{\left(\mathrm{wf}\right)}$ stand for
those from the baryon wavefunctions that also contains the linear 
$m_{\mathrm{s}}$. Expressions of transition magnetic moments of the
baryon antidecuplet $N_{\overline{\bm{10}}}$ to $N$ and those 
of the baryon eikosiheptaplet $N_{{\bm{27}}}\left(J=3/2\right)$: 
\begin{align}
\mu_{N N_{\overline{\bm{10}}}}^{(0)}
&=
-\frac{1}{6\sqrt{5}} \left(Q-1\right)
\left( w_1+w_2+\frac{1}{2}w_3\right),
\cr
\mu_{N N_{\overline{\bm{10}}}}^{(\mathrm{op})}
&=
-\frac{1}{54\sqrt{5}} \left(Q+1\right)w_4
-\frac{1}{18\sqrt{5}} \left(2Q-1\right)\left(w_5+\frac{3}{2}w_6\right),
\cr
\mu_{NN_{\overline{\bm{10}}}}^{\left(\mathrm{wf}\right)} 
& =  
-\frac{5}{24\sqrt{5}}\,c_{\overline{\bm{10}}} \,Q \left(
  w_1+\frac{5}{2}w_2-\frac{1}{2}w_3\right) 
\;-\;\frac{49}{72\sqrt{5}}\,c_{\bm{27}}\, \left(7Q-2\right) \left(
  w_1-\frac{11}{14}w_2-\frac{3}{14}w_3\right) 
\cr
&+\frac{1}{12\sqrt{5}}\,d_{\bm{8}}\left[2 \left(3-7Q\right)
  \left(w_1-\frac{1}{2}w_2\right)+ \left(Q+1\right)w_3\right] 
\cr
&-\frac{1}{90\sqrt{5}}\,d_{\bm{27}}\left(Q+1\right)
   \left(w_1+2w_2-\frac{3}{2}w_3\right).  
\label{eq:muN10bar}
\end{align}
The following are the expressions of the magnetic transition moments
of the baryon eikosiheptaplet $N_{{\bm{27}}}\left(J=3/2\right)$  to
the baryon octet:  
\begin{align}
\mu_{NN_{{\bm{27}}}}^{\left(0\right)} 
& =  
\frac{2}{9\sqrt{30}}\left(Q+1\right)\left(w_{1}+\frac{1}{2}w_{2}\right),
\cr
\mu_{NN_{{\bm{27}}}}^{\left(\mathrm{op}\right)} 
& =  
\frac{1}{126\sqrt{30}}\left(4Q-17\right)\left(w_{4}-\frac{11}{13}w_{5}+\frac{7}{13}w_{6}\right),
\cr
\mu_{NN_{{\bm{27}}}}^{\left(\mathrm{wf}\right)} 
& =  
-\frac{35}{18\sqrt{30}}\,  c_{\overline{\bm{10}}}(7 Q-2) \left(w_1-w_2\right) 
+\frac{1}{14 \sqrt{30}}\,c_{\bm{27}}\left[(19 Q-16) w_1 - 8 (Q+1) w_2\right].
\label{eq:muN27exp}
\end{align}
\end{appendix}

\end{document}